\documentclass{article}

\usepackage{arxiv}

\usepackage[utf8]{inputenc} 
\usepackage[T1]{fontenc}    
\usepackage{hyperref}       
\usepackage{url}            
\usepackage{booktabs}       
\usepackage{amsfonts}       
\usepackage{nicefrac}       
\usepackage{microtype}      
\usepackage{graphicx}

\title{Universality of population recovery patterns after disasters}

\author{
  Takahiro Yabe \\
  Lyles School of Civil Engineering\\
  Purdue University, USA\\
  \texttt{tyabe@purdue.edu} \\
   \And
  Kota Tsubouchi \\
  Yahoo Japan Corporation\\
  Tokyo, Japan\\
  \texttt{ktsubouc@yahoo-corp.jp} \\
 \And
  Naoya Fujiwara \\
  School of Information Sciences\\
  Tohoku University, Japan\\
  \texttt{fujiwara@se.is.tohoku.ac.jp} \\ 
  \And
  Yoshihide Sekimoto \\
  Institute of Industrial Science\\
  Tokyo, Japan\\
  \texttt{sekimoto@iis.u-tokyo.ac.jp} \\
   \And
  Satish V. Ukkusuri \\
  Lyles School of Civil Engineering\\
  Purdue University, USA\\
  \texttt{sukkusur@purdue.edu}
}

\begin{document}
\maketitle

\begin{abstract}
Despite the rising importance of enhancing community resilience to disasters, our understanding on how communities recover from catastrophic events is limited. Here we study the population recovery dynamics of disaster affected regions by observing the movements of over 2.5 million mobile phone users across three countries before, during and after five major disasters. We find that, although the regions affected by the five disasters have significant differences in socio-economic characteristics, we observe a universal recovery pattern where displaced populations return in an exponential manner after all disasters. Moreover, the heterogeneity in initial and long-term displacement rates across communities across the three countries were explained by a set of key universal factors including the community’s median income level, population size, housing damage rate, and the connectedness to other cities. These universal properties of recovery dynamics extracted from large scale evidence could impact efforts on urban resilience and sustainability across various disciplines.
\end{abstract}


Following the series of natural hazards with unprecedented severity and magnitude including Tohoku Tsunami and Hurricanes Harvey, Irma and Maria, the concept of urban resilience has gained significant global attention \cite{kull2016building,gitay2013building}. 
For many cities, it is of utmost importance to build institutional and infrastructural capacities to minimize economic loss and maintain the wellbeing of their citizens \cite{eakin2017opinion,kousky2014informing}. 
Recent disasters have shown the existence of large variance and disparities in recovery trajectories across cities and communities that experienced similar damage levels \cite{finch2010disaster,aldrich2012building}. 
We have witnessed manifold cases where cities experience significant drainage of population even after sufficient recovery of infrastructure systems \cite{myers2008social,mccaughey2018socio}. 
Understanding the interplay between the recovery of infrastructure systems and displaced populations after such large-scale disasters is essential for enhancing effective population recovery in communities, and to foster sustainable development in hazard prone areas \cite{aerts2018integrating}.  

City-level human migration patterns have long been studied based on surveys and census data \cite{sekimoto}, some of which focus on post-disaster migration \cite{mccaughey2018socio,cutter2008place,gray2012natural,finch2010disaster,fussell2014recovery,dewaard2016population,sadri2018role,piguet2011migration}. 
Such works reveal factors that are associated with migration, however often neglects the temporal patterns and spatial disparities of recovery and fails to provide general insights driven from multiple disaster instances. 
With the increase in the availability of large mobility datasets including mobile phone call detail records (CDR), GPS logs, and social media posts, longitudinal observations of individual mobility have become possible \cite{deville2014dynamic,jiang2016timegeo,wardrop2018spatially,gonzalez2008understanding,schneider}. 
A number of studies have focused on analyzing human mobility during and after earthquakes \cite{lu2012predictability,song2014prediction,yabe2016framework,yabe2018cross}, hurricanes \cite{wang2016patterns,wang2014quantifying}, and other anomalous events \cite{gao2016universal,bagrow2011collective}. 
Despite such progress, such studies are fragmented and there is no general understanding on the population displacement and recovery patterns of communities across heterogeneous disasters and countries.

In order to bridge this gap in the current literature, we analyzed large scale mobile phone GPS dataset from multiple disasters to unravel the general patterns of recovery after different disasters across different countries and to identify key mechanisms that explain common factors that explain such recovery patterns. 
More specifically, we collaborated with 3 different companies across US and Japan that collect GPS location data from mobile phones, and study the movements of more than 2.5 million mobile phones of affected individuals over a six-month period. 
We study five disasters that in total destroyed more than 1.5 million households, caused power outages in more than 8 million households, and caused more than \$350 billion in economic loss. 
The disasters studied are Hurricane Maria (Puerto Rico, USA, 2017), Hurricane Irma (Florida, USA, 2017), Tohoku Tsunami (Tohoku area, Japan, 2011), Kumamoto Earthquake (Kyushu area, 2016), and Kinugawa Flood (Ibaraki area, Japan, 2015), shown in \textbf{Figure 1a}. 
The collection of disasters in this study are heterogeneous in various aspects, including the type of disaster, location of occurrence, and socio-economic characteristics of communities in the affected regions. 

\section*{Results}
\subsection*{Universal Patterns of Population Recovery}
To observe the longitudinal displacement and recovery of population after the five disasters, we analyze the individual mobility patterns of over 1.9 million individuals who were identified to be living in the affected areas before the disaster. 
Each individual's home location was estimated applying a weighted mean-shift clustering on the GPS location points observed during nighttime \cite{ashbrook2003using,kanasugi2013spatiotemporal}. 
The rate of displacement on a given day after a disaster is defined as the rate of individuals that are staying away from their residential area.

Despite the differences in the type of disaster and distinct socio-economic characteristics of the affected regions (i.e. Puerto Rico, Florida and Tohoku area), the recovery patterns after the five disaster cases can all be well approximated by a negative exponential recovery pattern: $D(t)= (D_{LT}-D_0) \exp{(-\frac{t}{\tau})} + D_{LT}$, where $D_{0}$ is the initial displacement rate, $D_{LT}$ is the long term displacement rate, and $\tau$ is the recovery speed (\textbf{Figure 1b-f}).
Minor fluctuations observed in the recovery patterns are due to national holidays such as Christmas at around day 100 of Hurricane Maria and day 110 of Hurricane Irma, Thanksgiving Day at around day 80 of Hurricane Irma, and ``Obon Breaks'', which is a national holiday in Japan, at around day 120 of Kumamoto Earthquake. 
This negative exponential pattern suggests that the majority of people returns in the early days after the disaster, with the largest fraction returning on the first day in all disasters.
Stability of long-term displacements, implying the permanent migration of individuals from disaster affected areas, is another universal characteristic of recovery patterns across disasters. 
Our analysis shows that observing the displacement rates after the initial 50 days after the disaster is sufficient to predict the long-term displacement rate, implying the importance of efforts to decrease the initial displacement rates in the early stages after the disaster. 

\begin{figure}
\centering
\begin{minipage}{0.38\textwidth}
    \centering
  \includegraphics[width=\textwidth]{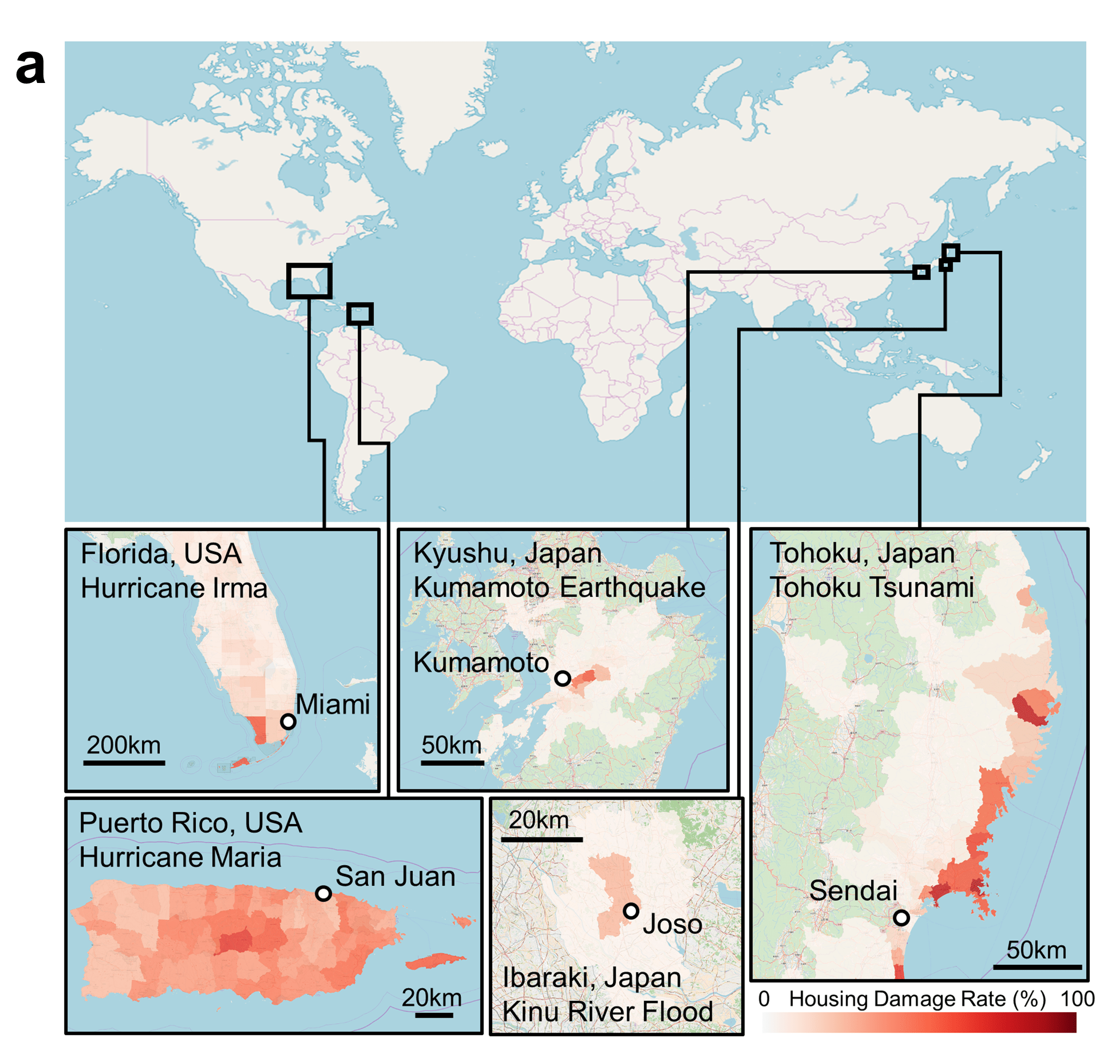}
  \label{fig:locations}
  \end{minipage}
\begin{minipage}{0.61\textwidth}
\centering
\includegraphics[width=\textwidth]{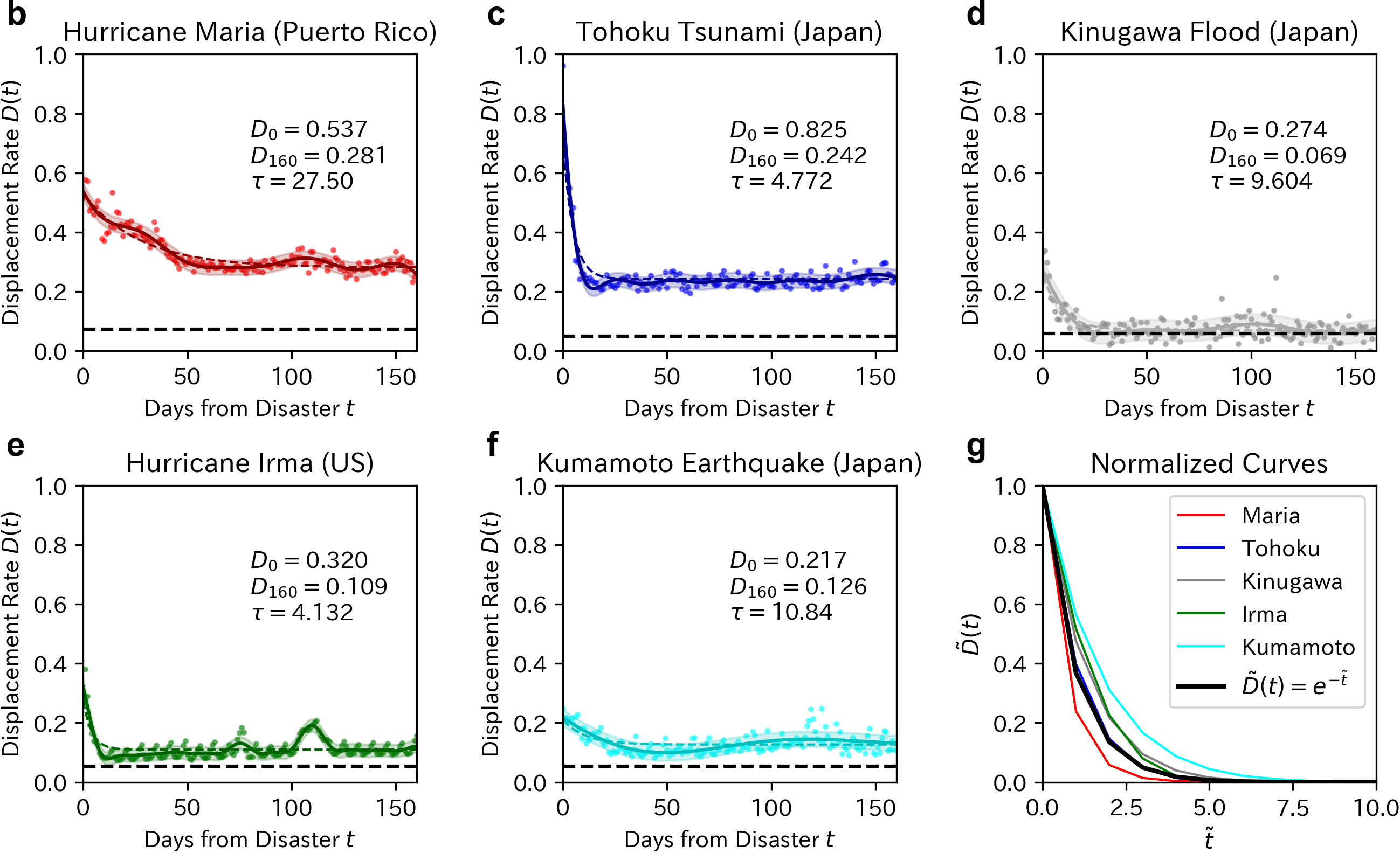}
\label{fig:ps}
\end{minipage} 
\caption{
\textbf{Universality of recovery patterns across various disasters.}  
\textbf{a.} 
Location, spatial scale, and severity of all disasters that were studied in this work. Red color indicates the percentages of houses that were severely damaged in each community in all disasters.
\textbf{b-f.} 
Population displacement and recovery patterns after each disaster. Raw observations of displacement rates are denoised using Gaussian Process Regression and are also fitted with a negative exponential function. Initial displacement $D_0$, long term displacement rate $D_{160}$ and recovery time $\tau$ are parameters of negative exponential function when fitted to raw data. Black horizontal dashed line shows average displacement rates on usual days before disaster.
\textbf{g.} 
Normalized displacement and recovery patterns after Hurricane Maria (red), Tohoku Tsunami Tsunami (blue), Hurricane Irma (green), Kumamoto Earthquake (cyan), Kinugawa Flood (gray), and $\tilde{D}(t)= e^{-\tilde{t}}$ (black) as reference. 
}
\label{fig:univpatterns}
\end{figure}

\begin{figure}
\centering
\includegraphics[width=0.8\textwidth]{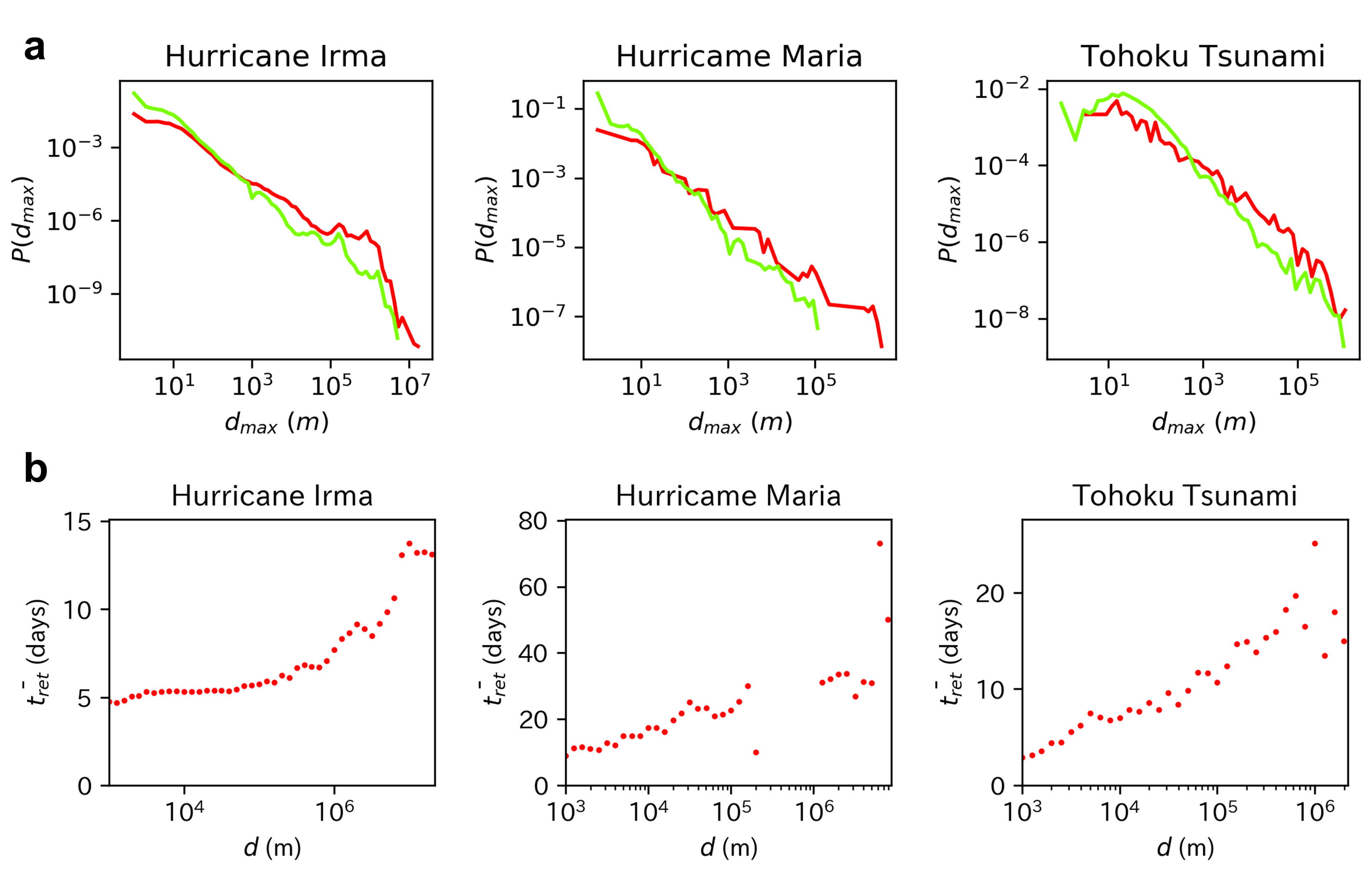}\label{fig:threed}
  \caption{
  \textbf{Distance and duration of evacuation.}
  \textbf{a.} 
   Probability distribution of maximum distance from home on a usual day (green) and 3 days after the disaster day (red). For all major disasters, we see an increase in populations staying away from their home locations more compared to usual times, reflecting the evacuation movements. It is also worth noting that distance distributions follow a power law both in usual times and emergency times for all disasters. 
   \textbf{b.} 
   Average evacuation duration plotted against evacuation distance. For all three major disasters (Hurricanes Irma, Maria and Tohoku Tsunami), we see an increase in evacuation duration for individuals that evacuated for a longer distance. This phenomenon partially explains why people return in an exponential manner, where majority of the people who evacuated to nearby locations return right after the disasters, and the people who evacuated to locations far away take more time to return back to their homes.
  }
  \label{fig:thisfig}
\end{figure}

The difference in parameter values across disasters can be explained by the infrastructure recovery speed in each of the affected regions. The recovery time $\tau$ is relatively short ($4 < \tau < 10$) in disasters that occurred in Japan and mainland US, while very large $\tau=27.5$ for Hurricane Maria that occurred in Puerto Rico. 
The recovery time needed for physical infrastructure (e.g. power, water, gas, transportation\cite{rinaldi2001identifying}) had similar trends, where in Japan and mainland USA, power was restored in over 90\% of the households (that were not destroyed) within 10 days from the disaster, while it took more than 200 days for Puerto Rico (\textbf{Figure S1}).
To show that all observations can be well explained by a negative exponential function, the observations of displacement rates $D(t)$ were normalized for all disasters (\textbf{Methods}). 
\textbf{Figure 1g} shows the normalized displacement rate observations $\tilde{D}(\tilde{t})$ for each disaster in colors, along with the negative exponential function ($\tilde{D}(\tilde{t}) = e^{-\tilde{t}}$) shown in black.
The closeness in the trends of the normalized displacement rates with the negative exponential function shows that for all disasters, population recovery curves can be well explained by a negative exponential function.

The negative exponential population recovery pattern which was observed across different disasters can be explained from the observations of individual mobility patterns after disasters.
\textbf{Figure 2a} shows the probability distributions of the maximum distance traveled from estimated home locations on before and post disaster timings of users who were affected by the disaster, in green and red, respectively. 
To compute the distribution of maximum traveled distances before the disaster, locations of individuals 3 days before the disaster was used for all disasters. 
Similarly, the distributions of post-disaster maximum distances were computed by analyzing the movements of the affected individuals 3 days after each disaster. 
We can clearly see for all disasters that the probability of people staying closer to their homes decrease after the disaster compared to usual days, and that people with long distances increase after disasters, indicating evacuation activities. 
The differences between each curve highlights how far people evacuate to in each disaster. 
We observe a large population evacuating around 100 to 1000 kilometers after disasters in the US, while people in Japan evacuate around 1 to 100 kilometers, which roughly reflects the sizes of the countries. 
\textbf{Figure 2b} shows the average evacuation duration against the evacuated distance. We can observe that the evacuation duration increases as evacuation distances increase in all three major disasters. 
This is assumed to be because the cost of transportation for returning back home increases as the distance becomes longer.
Thus, combined with the findings from \textbf{Figure 2a}, we can conclude that after disasters, the majority of the people who evacuated to nearby close locations return right after the disaster, while the minority people who evacuated to locations far away took more time until their return, in aggregate exhibiting a negative exponential recovery pattern.

\subsection*{Estimating Population Recovery Outcomes}
\textbf{Figures 3a-3d} show the initial displacement rates $D_0$ plotted against housing damage rates $r_h$ for each community in the four major disasters. 
Solid lines are regression results using least squares method and the dotted black lines mark theoretical results if all people whose houses were damaged fled their hometown. Pearson's R values are shown in the corner of each panel, and their significance levels are noted by stars (*) beside the Pearson's R (***: $p<0.01$). 
For disasters in Japan and mainland US, housing damage and initial displacement rates had a significant and strong correlation, indicating that physical damage directly affects the evacuation decisions of people.
On the other hand, the correlation was insignificant and less in Puerto Rico after Hurricane Maria. Rather, as shown in \textbf{Figures 3e-3g}, median income levels and number of households for each community had significant and stronger correlations with initial displacement rates. 
The Figure shows the spatial distribution of initial displacement rates $D_0$ (Figure 3e) median income $MI$ (Figure 3f), and the number of households $N_h$ (Figure 3g) for all communities in Puerto Rico after Hurricane Maria. 
Median income and number of households had negative correlations with initial displacement rates, indicating that communities with lower incomes and smaller populations had higher initial displacement rates. 
This indicates that in countries that have low income levels and poor standards for infrastructure, the wealth of communities governs how communities experience initial displacement rates. 
However, for all disasters, the correlation between housing damage rates and long-term displacement rates were weaker than that with initial displacement rates (\textbf{Figure S2}), implying that recovery of communities are governed by more factors in addition to housing damage rates.

\begin{figure}
\centering
\includegraphics[width=\textwidth]{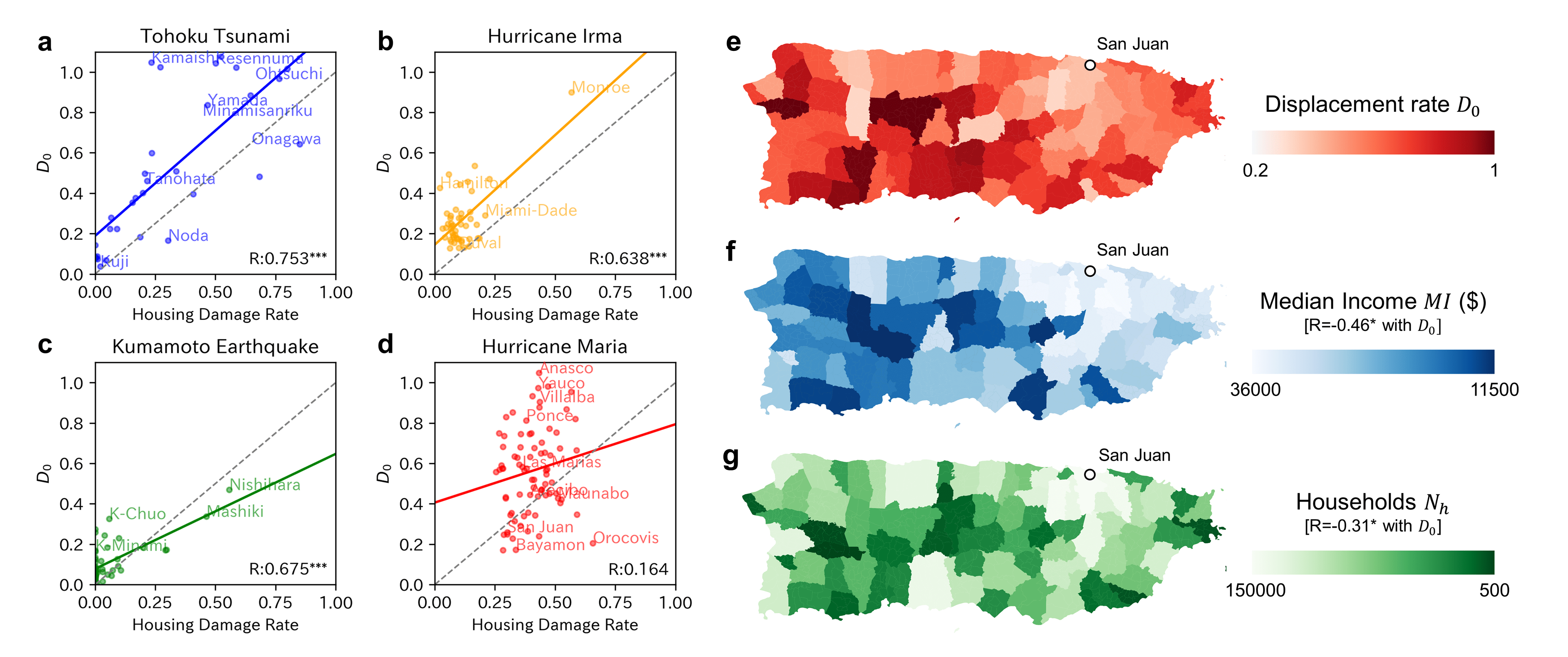}
\label{fig:threed}
  \caption{
  \textbf{Initial displacement rates after disasters.}
  \textbf{a-d.} 
  Initial displacement rates $D_0$ plotted against housing damage rates for each community for Tohoku Tsunami (a), Hurricane Irma (b), Kumamoto Earthquake (c), and Hurricane Maria (d). 
  Solid lines are regression results and the dotted black lines mark theoretical results if all people whose houses were damaged fled their hometown. 
  Pearson's R values are shown in the corner of each panel, and their significance levels are noted by stars (***) beside the Pearson's R (***: $p<0.01$). 
  For disasters in Japan and mainland US, housing damage and initial displacement rates had a significant and strong correlation, indicating that physical damage directly affects the evacuation decisions of people. 
  On the other hand, the correlation was insignificant and less in Puerto Rico after Hurricane Maria. 
  \textbf{e-g.} Spatial distribution of displacement rates (e), median income (f) and number of households (g) for all communities in Puerto Rico after Hurricane Maria. 
  For Hurricane Maria, median income levels and number of households for each community had significant and negative correlations with initial displacement rates, indicating that communities with lower incomes and smaller populations had higher initial displacement rates.
  }
  \label{fig:thisfig}
\end{figure}

\begin{figure}
\centering
\includegraphics[width=0.8\textwidth]{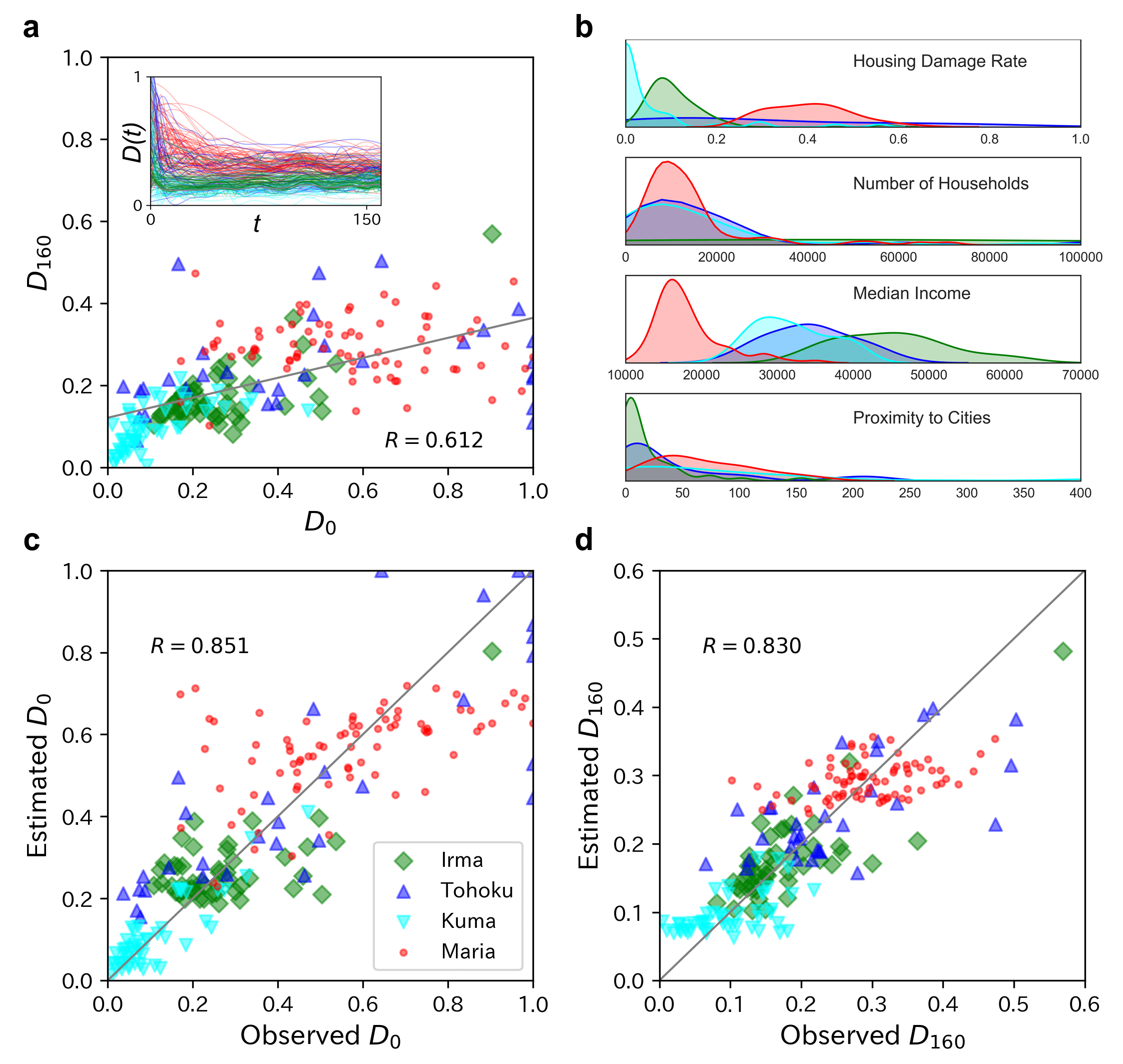}
\label{fig:threed}
  \caption{
  \textbf{Universal key factors explain community level recovery. }
  \textbf{a.} 
  Observed initial and long term displacement rates of all communities in the four major disasters. One plot corresponds to one community, and colors represent the disaster each community experienced. The initial and long term displacement rates have a moderate correlation of $R=0.612$. The inset shows the recovery trajectories of all communities, which show high heterogeneity in each disaster. 
  \textbf{b.} Kernel density plots of the four attributes of the communities that were affected by each disaster, showing the heterogeneity in socio-economic characteristics of communities across and within the four disasters. 
  \textbf{c.} 
  Observed and estimated initial displacement rates for all communities in the four major disasters. Displacement rates were estimated using the linear combination of the four key factors: housing damage rate $r_h$, number of households $N_h$, median income $MI$, and proximity to cities $d_p$. $R=0.851$ is the correlation coefficient between observed and estimated values.
  \textbf{d.} Estimated long term displacement rates on day 160 from each disaster, compared with observed values with correlation $R=0.830$.
  }
  \label{fig:thisfig}
\end{figure}

To further investigate whether universal characteristics can be observed across disasters, we downscale our analysis to communities (counties in the US, wards/cities in Japan) within each region affected by the five disasters. 
In total, we observe recovery patterns of 263 communities with large diversity in socio-economic characteristics including population size and income levels. 
The inset in \textbf{Figure 4a} shows the large heterogeneity in recovery patterns across communities, even within each disaster. 
The main plot in \textbf{Figure 4a} shows the relationship between initial and long-term displacement rates for all communities in the four major disasters. 
Initial displacement and long-term displacement had moderate level of correlation (R=0.612), which implies that communities recover in heterogeneous ways. 
To identify the factors that govern the communities' recovery patterns, we performed ANOVA on displacement rates observed in each community. 
Factors including housing damage rates $r_h$, number of households $N_h$, household median income $MI$, proximity to large cities $d_p$, and infrastructure recovery time $t_R$ were used in the analysis.
$d_p$ measures the amount of surrounding population for a given city with respect to its own size (\textbf{Methods}). 
These variables were shown to have low correlations among themselves (\textbf{Table S1}). 
Infrastructure recovery time $t_R$ was not included in the model for estimating initial displacement rates, since this information would not be available at time t=0. 
The probability density of the four attributes in each disaster are shown in \textbf{Figure 4b}. 
We can see a clear difference in housing damage rates and median income levels across the four disasters, but more similarity in the number of households per community and the proximity of cities. 
Despite the distinct characteristics of communities and disasters, we show that this set of universal factors can explain displacement and recovery performances of communities. 
\textbf{Figure 4c} shows the estimation performance of the linear regression results for initial displacement rates $D_0$. 
Although we utilize only four variables for estimation, the estimated displacement rates have a high correlation with observed values (R=0.851). 
For disasters in Japan (Tohoku Tsunami and Kumamoto Earthquake) and US (Hurricane Irma), the housing damage rate was significant in all cases.
However, Puerto Rico (Hurricane Maria) does not have a significant correlation with housing damage rates, but with median income, proximity to other cities, and the number of households. 
All models were statistically significant at p-value of 0.01. 
Similarly, \textbf{Figure 4d} shows that long term displacements $D_{160}$ were explained by the five key variables with high accuracy (R=0.830).
Despite the large heterogeneity in community features and disaster characteristics across the four major disasters, these key factors sufficiently explained the variance in displacement rates at different time points, implying the universality of these factors in understanding the recovery of communities. 
For all disasters, the housing damage rate was significant in all cases, indicating that long term recovery does depend on how much direct damage the community suffers. 
All models were statistically significant at p-value of 0.1. Interestingly, the number of households, median income, proximity to cities, and infrastructure recovery time were each significant for different disasters, showing heterogeneity in how these variables affect recovery processes in in the four disasters. 
Detailed regression results are shown in \textbf{Table S2} and \textbf{Table S3}.

To increase the fairness in the comparison of the recovery outcomes across communities, we analyze the heterogeneity in recovery patterns conditional to initial displacement rates. 
All of the community recovery patterns from all five disasters were divided into 3 groups based on similarities of initial displacement rates ($0.8<D_0 \leq 1.0$, $0.6<D_0 \leq 0.8$, $0.4<D_0 \leq 0.6$) (\textbf{Figure S3}). 
By mixing the recovery patterns from all of the disasters based on similar initial displacement values, we attempt to assess whether we can observe similar community recovery patterns across different disasters, and to extract general findings on the factors that govern recovery patterns. 
Hierarchical clustering was performed to further classify the communities into clusters of speedy, moderate and slow recovery (\textbf{Figure S4}). 
It is found that communities across countries with different characteristics may resemble similar recovery patterns, especially in the group with high initial displacement values. 
More specifically, some communities affected by Tohoku Tsunami show similar patterns with communities affected by Hurricane Maria, despite large differences in socio-economic characteristics and infrastructure recovery speeds. 
In contrast, communities with lower initial displacement values tend to differ by disasters, where communities in Puerto Rico tend to have slower recovery and communities in Florida and Tohoku have quicker recovery. 
Moreover, the timing of maximum variance across communities was earlier for communities with high initial displacements compared to communities with lower initial displacements (\textbf{Figure S5}). 
This result reveals that the decision-making mechanisms of individuals are conditional on the intensity of the initial shock. 
When damage levels are severe, returning decisions are made in the early stages (within a week) from the disaster based on early observations of the disaster damage. 
On the other hand, people tend to stay and observe the situations of their community for a longer period of time if the damage seems relatively moderate. 
An increase in displacements was observed in only Puerto Rican communities, which imply that more people started leaving after around 2 weeks due to slow recovery of utilities. 

ANOVA results on these 3 groups of mixed communities revealed that connectedness with other cities plays a crucial role in recovery performances in addition to socio-economic variables of communities, especially significant for communities with high initial displacements (\textbf{Table S4}). 
This finding contradicts previous findings from analysis on non-disaster human mobility patterns (e.g. commuting) that models the increase in migration probability according to the amount of opportunities available in surrounding cities \cite{simini2012universal}. 
Rather, in disaster situations, our analysis confirms that the long-term displacement is lower in the presence of surrounding cities with high populations, implying that the recovery of cities can be enhanced by the provision of help and support from neighboring cities. 
This extends the theories on the importance social capital and social support \cite{aldrich2012building,sadri2018role} to an intercity-scale. 
\textbf{Figure 5} shows pairwise comparisons after Hurricane Maria and Tohoku Tsunami where a pair of communities that share similar disaster levels and socio-economic characteristics (e.g. housing damage level, population, income levels, time needed for infrastructure recovery) but have different connectedness to neighboring cities have distinct recovery outcomes. 
The effect of inter-city connectivity on community recovery is an understudied dimension in urban resilience, and has significant implications on the planning of inter-city networks to enhance the resilience of communities.

\begin{figure}
\centering
\includegraphics[width=\textwidth]{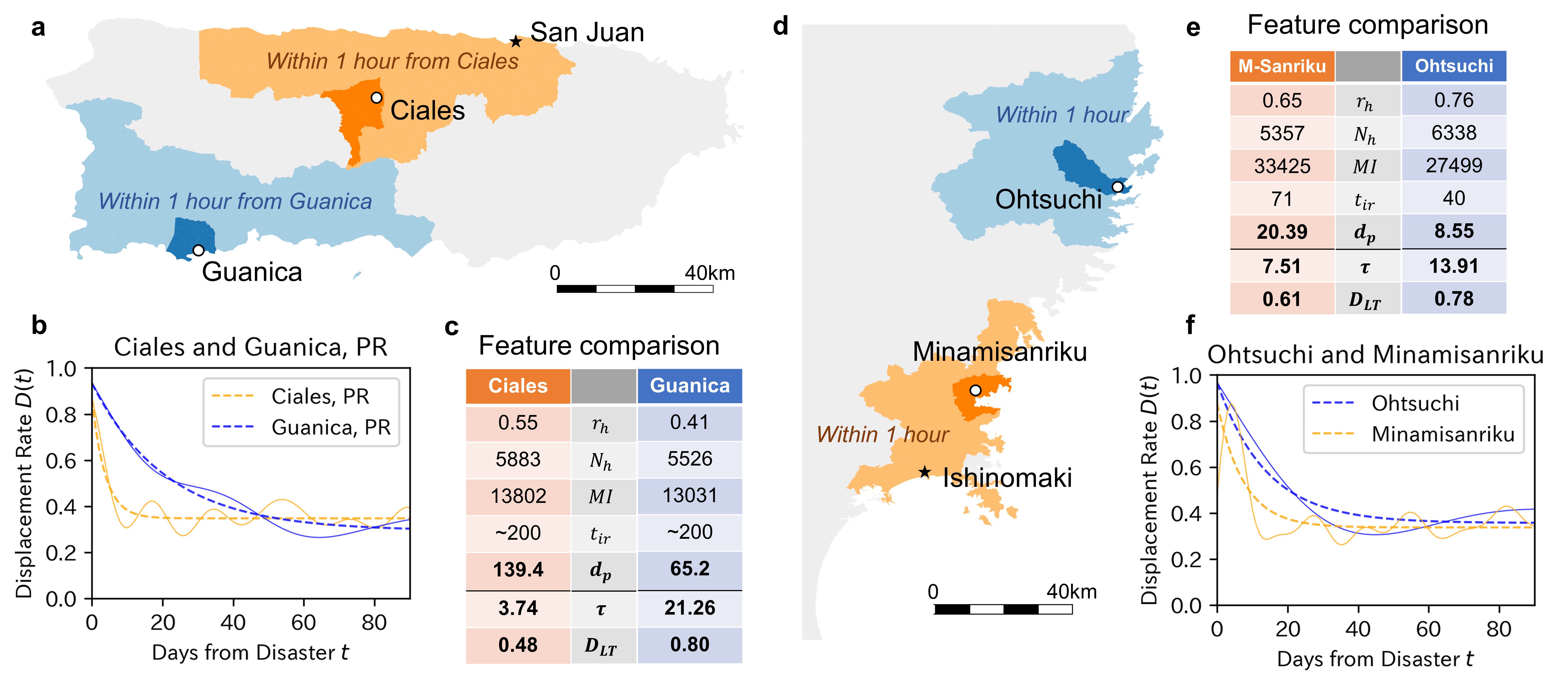}\label{fig:threed}
  \caption{
\textbf{Connectedness to neighboring cities as a key factor to recovery. }
  \textbf{a.} 
    Map of Ciales and Guanica, Puerto Rico. Light colored areas show the area that can be reached from each city within one hour of driving time. 
    \textbf{b.} Recovery patterns of both communities, showing the faster recovery of Ciales. 
    \textbf{c.} Comparison of factors for both cities. Factors other than connectedness to other large cities (e.g. San Juan) was similar between the two communities.
    \textbf{d-f.} 
    Similar phenomenon was seen after Tohoku Tsunami in Japan. Minamisanriku city and Ohtsuchi city shared similar characteristics except for the connectedness to large cities (e.g. Ishinomaki), resulting in differences in recovery speed. 
}
  \label{fig:thisfig}
\end{figure}

\subsection*{Discussion}
In this work, we used large scale mobile phone datasets from five major natural disasters across the US, Puerto Rico and Japan, to uncover the population recovery patterns. 
We found that population recovery patterns follow a universal negative exponential function in all of the disasters that we have analyzed. 
The majority of the population returns to their residential areas in the initial few days after the disaster, while some decide to permanently stay migrate to other cities. 
This extends the previous understanding of post-disaster human migration \cite{finch2010disaster,fussell2014recovery,dewaard2016population,sadri2018role} by providing a continuous and longitudinal understanding of the recovery dynamics. 
Further analysis on the individual evacuation mobility trajectories provided explanation on why such patterns emerge. 
Analysis of the distribution of evacuation distances confirms other studies \cite{lu2012predictability,yabe2018cross}, and extends their finding by further showing the correlation with evacuation duration. 
Such characteristics of individual mobility patterns after disasters can be applied in developing agent based simulations of evacuation and return mobility, which are commonly used in practice to predict post-disaster mobility and population recovery. 
Furthermore, the negative exponential pattern implies that the long-term displacement rate stays static after some transition period, which can help officials to predict the long-term population dynamics after disasters.
This finding also informs officials on the importance of early-stage policy implementations to influence more people to return back to their original residential areas. 

Through a cross comparative analysis of population displacement and recovery patterns of over 250 communities across different disasters in different countries, we showed that the variance in initial and long-term displacement rates can be well explained by key common factors, including community population, median income, housing damage rates, proximity to other cities, and infrastructure recovery time. 
These findings extend the insights obtained from individual case studies \cite{myers2008social,mccaughey2018socio,gray2012natural,finch2010disaster,fussell2014recovery} that were carried out in some disaster instances to a global and generalizable scale. 
On the other hand, some findings including the importance of proximity to other cities, introduced new perspectives on community recovery. 
The importance of physical proximity to other cities has large implications on policy making for disaster resilience. 
Instead of concentrating on the evaluation of each community in an individual manner, this finding suggests that policy makers need to evaluate the collective capacity of the network of communities to predict the recovery of communities. 
This insight would lead to interesting problems related to network design and infrastructure investment allocation. 
In the case of Puerto Rico after Hurricane Maria (\textbf{Figure 5a}), the arterial road that runs through the central part of the island connecting communities in the southern part to the northern cities (e.g. San Juan) is not well designed, with inefficient road structures and paths. 
Similarly, communities in the northern part of Tohoku region (\textbf{Figure 5d}) are severely isolated due to the mountainous terrain.
The results provided in this paper could raise the awareness of the importance of road network design for improving the resilience of cities.

Several limitations exist in this work. 
First, although several works have confirmed its high estimation accuracy when aggregated in the community scale \cite{yabe2016framework,song2014prediction}, they may contain demographic biases. The representativeness of cell phone data may dynamically vary across cities over time, making it technically infeasible to validate using surveys for every city with high temporal granularity.
Nevertheless, an integrative study using both large-scale cell phone location data and detailed household surveys is suggested for future studies. 
Second, although our study analyzed the longitudinal population recovery until 160 days after the disaster, a longer study could provide more insights in the recovery of communities. 
Especially for Puerto Rico after Hurricane Maria, observing more longitudinal data to determine whether the displacement rate permanently stays high after 160 days would provide valuable insights. 
With data for longer time periods from more instances of large-scale disasters in different countries, our findings will enable more accurate predictions and effective controls of population recovery after disasters.
Third, our primary focus of this paper was on the return mobility of the displaced populations after disasters. 
Extending this study to not only returning behavior but also incoming migration would be of interest to understand the recovery and further development of each community. 
Moreover, this study was limited to the analysis of correlations of certain variables to recovery outcomes. 
Using a more dynamic modeling approach and investigating in the underlying mechanisms of such recovery patterns could lead to better understandings of population recovery after disasters. 
However, we believe that this work lays out a foundation for studies on understanding the population recovery dynamics of disaster affected cities, which is an increasingly important problem for policy makers in urban planning and disaster management.

\section*{Methods}
\subsection*{Mobile Phone Location Data} Mobile phone location data for the six disasters were provided by 3 different companies in Japan and the US.
Location data were collected by Yahoo Japan Corporation \footnote{https://www.yahoo.co.jp/} for Kumamoto Earthquake and Kinugawa Flood, by Zenrin Data Com \footnote{http://www.zenrin-datacom.net/toppage} for Tohoku Tsunami and Earthquake, and Safegraph \footnote{https://www.safegraph.com/} for Hurricanes Irma and Maria. 
All companies obtained the location information (time, longitude, latitude) of mobile phones via the Global Positioning Satellite (GPS) system. 
GPS data were obtained from mobile phones of individuals who agreed to provide their location data for research purposes, and all information were anonymized to protect the security of users. 
For each disaster, the area of study was the set of communities that had housing damage of more than 15\%, which are shown in \textbf{Figure 1a}. Cell phone data for 6 months before and after the disaster date was observed for all disasters. Each user was observed at high frequency for all datasets, although there were differences in the average observation points per user per day. 
Although the average observations vary across datasets, we only observe where each user is staying overnight, which does not require that much of a high granular dataset.
\textbf{Table S1} shows the detailed information on the datasets.

\subsection*{Location Data Analysis}
It is well known that human trajectories show a high degree of temporal and spatial regularity, each individual having a significant probability to return to a few highly frequented locations, including his/her home location \cite{gonzalez2008understanding}.
Due to this characteristic, it has been shown that home locations of individuals can be detected with high accuracy by clustering the individual's stay point locations over night \cite{calabrese2011estimating}.
Home locations of each individual was detected by applying mean-shift clustering to the nighttime stay points (observed between 8PM and 6AM), weighted by the duration of stays in each location \cite{ashbrook2003using,kanasugi2013spatiotemporal}.
Mean shift clustering was implemented using the scikit-learn package on Python.
An individual was detected to be displaced if the individual is estimated to be staying in a location outside the city where his/her estimated home location belongs to. 
Displacement rate $D_c(t)$ on day $t$ for city $c$, is calculated by dividing the number of displaced individuals observed on day $t$ whose homes were estimated to be in city $c$, by the total number of individuals who were estimated to be residents of city $c$ observed on day $t$.

\subsection*{Gaussian Process Regression}
In \textbf{Figure 1b-f}, the raw observations of displacement rates were de-noised using Gaussian Process Regression. 
Gaussian Process Regression (GPR) is a non-parametric probabilistic model for denoising and regression \cite{rasmussen2006gaussian}.
Gaussian Processes (GPs) are extensions of multivariate Gaussian distributions to infinite dimensionality. 
GPs assume that values observed at $t=t_i$ and $t=t_j$ are jointly Gaussian with zero mean and covariance given by a covariance function $k(t_i, t_j)$. 
In this model, we use the squared exponential covariance function and we assume that the observed values $y$ have i.i.d. Gaussian noise with variance $\sigma_n^2$ added on, shown in the following equation: 
\begin{equation}
cov(f(t_i),f(t_j)) = k(t_i, t_j) = \exp \Big( {\frac{1}{2l^2} |t_i - t_j|^2} \Big) + \sigma_n^2 \delta_{i,j} 
\end{equation}
where $\delta_{ij}$ is a Kronecker delta which is 1 if $i=j$ and zero otherwise. 

In the GPR model, the hyperparameters are the length scale $l$ and the scale of the Gaussian noise of the observed values $\sigma_n$. 
The model chooses the hyperparameters and covariances directly from the training data. 
To obtain such optimal hyperparameters, the log marginal likelihood $L(\theta)$, shown below, is maximized with respect to hyperparameters and noise level $\theta = {l,\sigma_n}$.
\begin{equation}
L = \log p(y|t) = -\frac{1}{2}y^T(K+\sigma_n^2I)^{-1} - \frac{1}{2} \log |K+\sigma_n^2I| - \frac{n}{2} \log 2\pi
\end{equation}

The minimization of $L(\theta)$ is solved by conjugate gradients method \cite{ebden2008gaussian}.  
To implement the GPR model, we used the package available on scikit learn and implemented the model using Python codes\footnote{\url{http://scikit-learn.org/stable/modules/gaussian_process.html}}. 

\subsection*{Normalization of Recovery Curves}
In \textbf{Figure 1g}, recovery curves of each disaster were normalized to show that they all follow similar negative exponential patterns. 
Each curve was modeled by the negative exponential function:$D(t)= (D_{LT}-D_0) \exp{(-\frac{t}{\tau})} + D_{LT}$, where $D_{0}$ is the initial displacement rate, $D_{LT}$ is the long term displacement rate, and $\tau$ is the recovery speed.
Thus, if the empirical observations completely follows this law, all normalized curves of each disaster $\tilde{D}({t}) = \Big( \frac{D(t)-D_{LT}}{D_0 - D_{LT}} \Big)^\tau $ should collapse to the same curve $e^{-t}$.

\subsection*{Features for Regression}
The proximity of city $i$ to other cities is calculated by $d_p(i)= \frac{\sum_{j\in S(i)} N_j}{N_i}$, where $N_i$ is the number of households in city $i$, and $S(i)$ is the set of cities that can be reached within 1 hour by vehicles from city $i$. 
$d_{p}$ would be large for small cities that have large cities around it, and small for more isolated cities.
For cities with similar population levels, $d_p$ would be proportional to the total population of surrounding cities. 
All variables were normalized when regression was performed, by $\tilde{z}= \frac{z - \min(Z)}{\max(Z)-\min(Z)}$ where $\min(Z)$ and $\max(Z)$ are the minimum and maximum values of the variable, respectively.

\subsection*{Code Availability}
Computer codes used to process and analyze the data are posted on the author's github page (\url{https://github.com/takayabe0505}).

\section*{Acknowledgements} We thank Mr. Noah Yonack of Safegraph, Mr. Hodaka Kaneta of Zenrin Data Com, and Mr. Hiroshi Kanasugi of University of Tokyo for preparing the mobile phone GPS data used in this study.
The work of T.Y. and S. V. U. is partly funded by NSF Grant No. 1638311 CRISP Type 2/Collaborative Research: Critical Transitions in the Resilience and Recovery of Interdependent Social and Physical Networks. N.F. was supported by JSPS KAKENHI Grant Number JP17H01742.

\section*{Competing Interests} The authors declare that they have no
competing financial interests.

\section*{Correspondence} Correspondence should be addressed to S.V.U. (email: sukkusur@purdue.edu).

\section*{Author Contributions} T.Y., K.T., N.F., Y.S., and S.V.U. designed research; T.Y., K.T., N.F., and S.V.U. performed research; T.Y. and K.T. analyzed data; T.Y., K.T., N.F., Y.S., and S.V.U. wrote the paper.

\section*{Data Availability}
Mobile phone location data are proprietary data owned by private companies.
Although such data are not available for open access due to the users' privacy, we will obtain permission to post processed data that are sufficient to reproduce the results obtained in this study.
Data collected from other sources are available from official documents that are openly accessible.

\bibliographystyle{unsrt}  
\bibliography{references}  






\section*{Supplementary Material}

\newcommand{\hbAppendixPrefix}{S}
\renewcommand{\thefigure}{\hbAppendixPrefix\arabic{figure}}
\renewcommand{\thetable}{\hbAppendixPrefix\arabic{table}} 

This supplementary information contains the following:
\begin{description}
    \item Fig. S1. Recovery speed of power outages
    \item Fig. S2. Long term displacement rates and housing damage.
    \item Fig. S3. 3 groups of community recovery patterns.
    \item Fig. S4. Clustering results of community recovery patterns.
    \item Fig. S5. Temporal variance across communities.
    \item Table S1. Title of the first supplementary table.
    \item Table S2. Correlation between variables for all disasters.
    \item Table S3. Regression results for initial displacement rates.
    \item Table S4. Regression results for long term displacement rates.
    \item Table S5. Regression results for long term displacement rates for groups with different initial displacement values.
\end{description}

\setcounter{figure}{0}
\setcounter{table}{0}

\begin{figure}[p]
\begin{minipage}[c][\textheight]{\textwidth}
\centering
\includegraphics[width=0.75\textwidth]{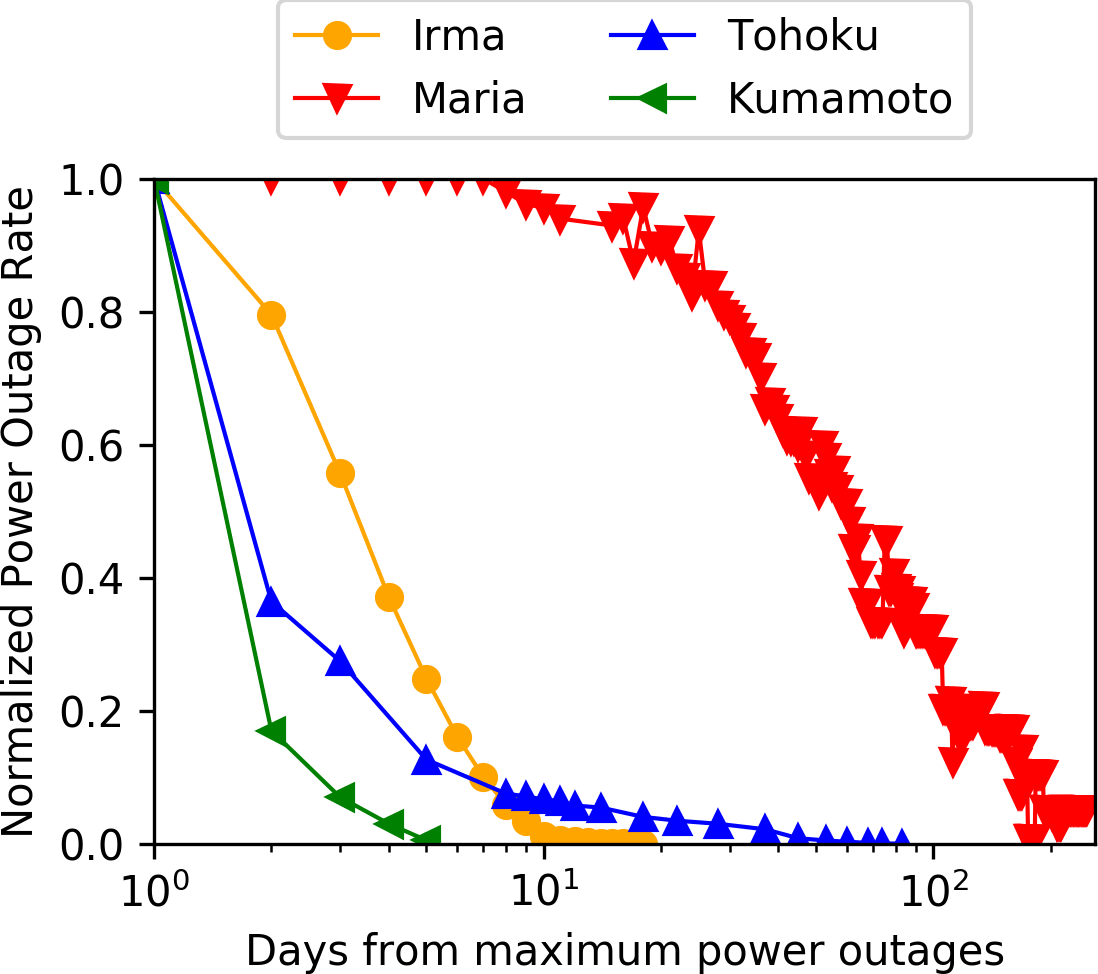}
\caption{\textbf{Recovery speed of power outages.} Comparison of recovery from power outages across the four major disasters. For Hurricane Irma (Florida, USA), Tohoku Tsunami (Tohoku, Japan), and Kumamoto Earthquake (Kyushu, Japan), it took less than 10 days from the disaster day to restore more than 90\% if the power outages. On the other hand, it took more than 200 days after Hurricane Maria for full recovery of power in Puerto Rico.}
\label{fig:poweroutage}
\end{minipage}
\end{figure}

\begin{figure}[p]
\begin{minipage}[c][\textheight]{\textwidth}
\centering
\includegraphics[width=\textwidth]{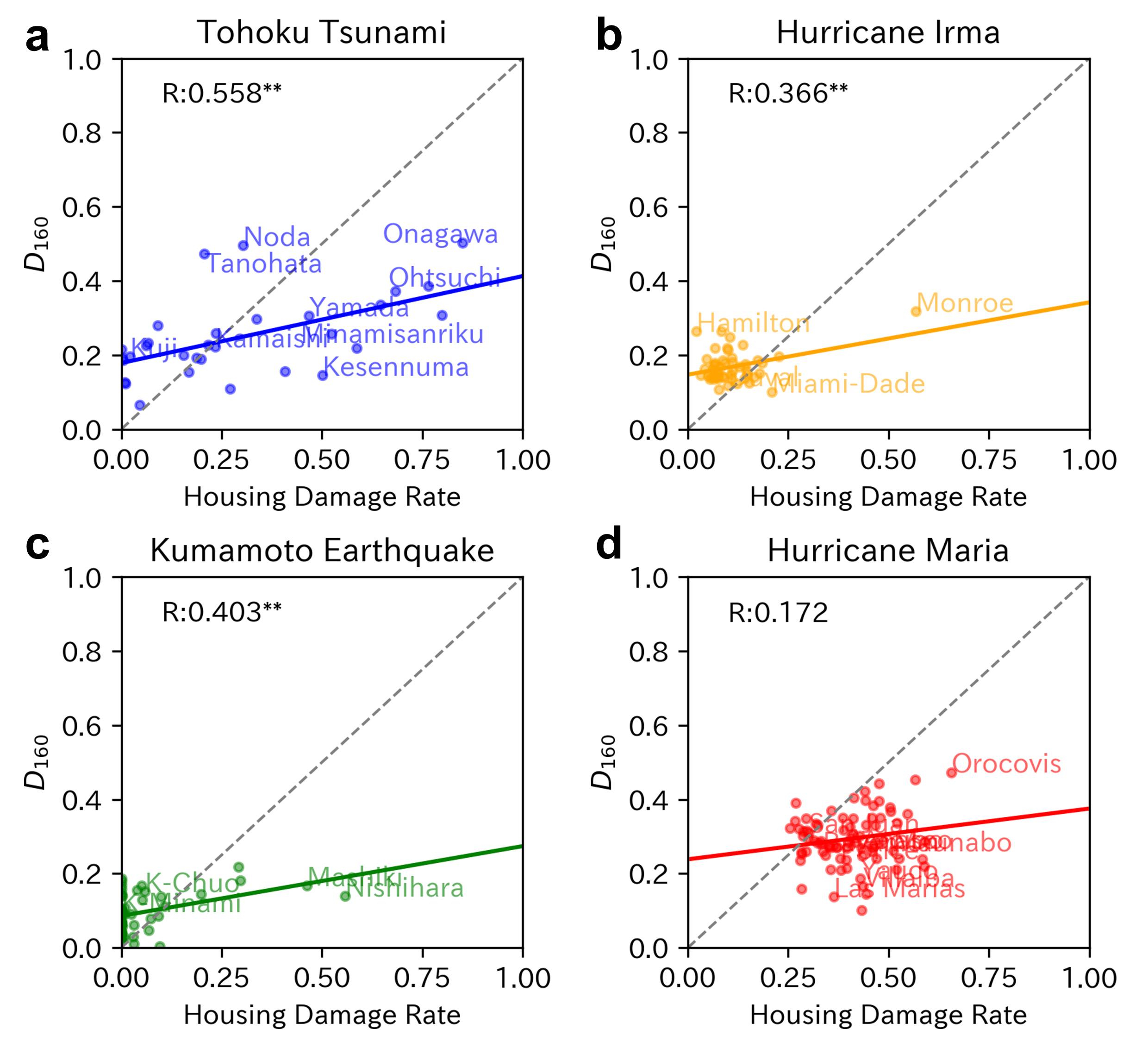}
\caption{\textbf{Long term displacement rates $D_{160}$ plotted against housing damage rates for each community in the four major disasters.}
For all disasters, the correlation between housing damage rates and long term displacement rates are weaker than that with initial displacement rates, implying that recovery of communities are governed by more factors in addition to housing damage rates. 
In some communities such as Noda and Tanohata after the Tohoku Tsunami (a), displacement rates even increased compared to initial displacement rates after the Tohoku Tsunami. 
}
\label{fig:lt}
\end{minipage}
\end{figure}

\begin{figure}[p]
\begin{minipage}[c][\textheight]{\textwidth}
\centering
\includegraphics[width=0.6\textwidth]{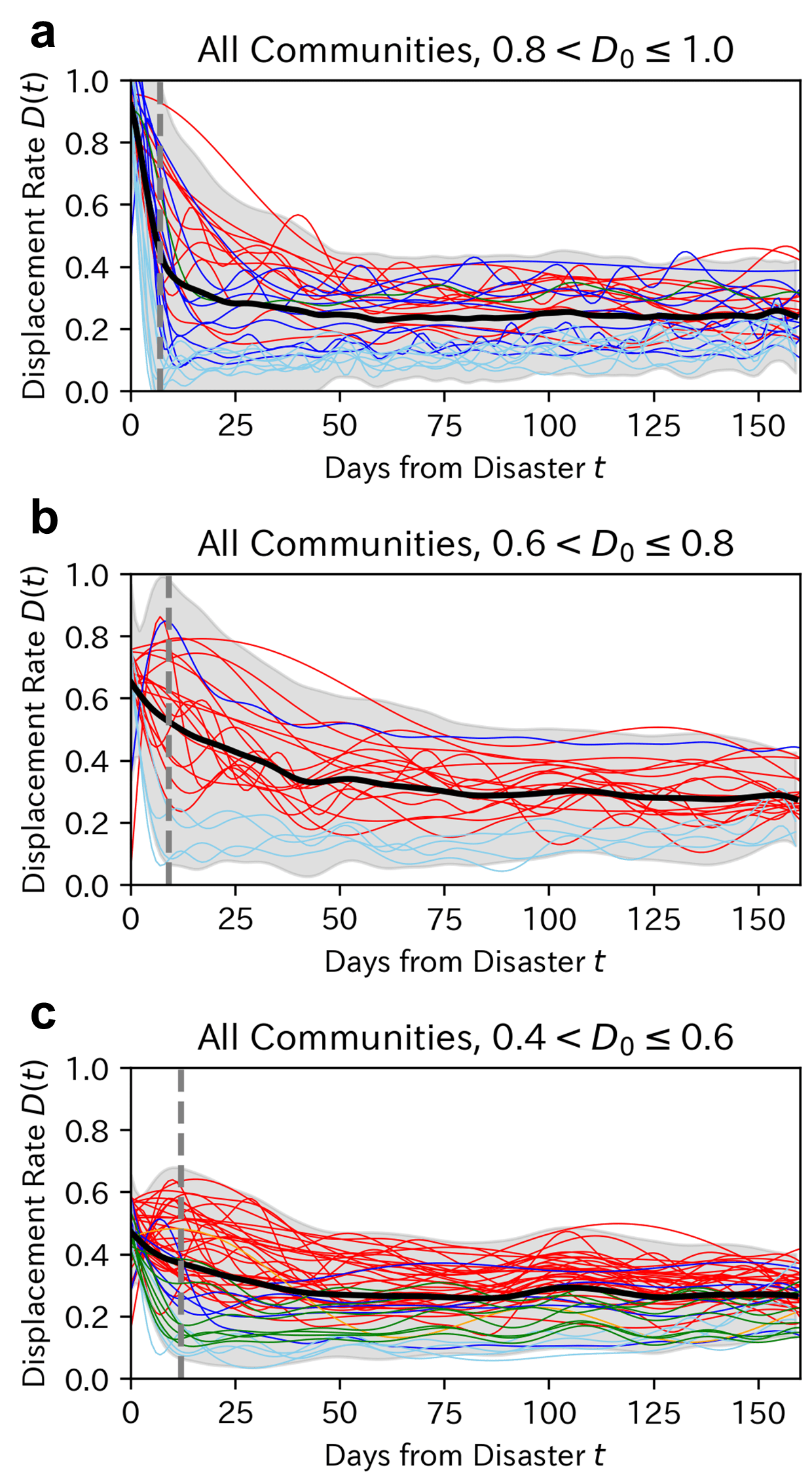}
\caption{\textbf{3 groups of community recovery patterns.} Recovery patterns $D(t)$ of each community plotted against number of days after the disaster for all disasters. Community recovery patterns are divided into three groups depending on the initial displacement rates. Colors of the patterns denote the disaster each community experienced, similar to previous figures. Broad black line shows the average values of the recovery patterns in the group. The results show that long term recovery outcomes are heterogeneous despite similar initial displacement conditions. 
}
\label{fig:3groups}
\end{minipage}
\end{figure}

\begin{figure}[p]
\begin{minipage}[c][\textheight]{\textwidth}
\centering
\includegraphics[width=\textwidth]{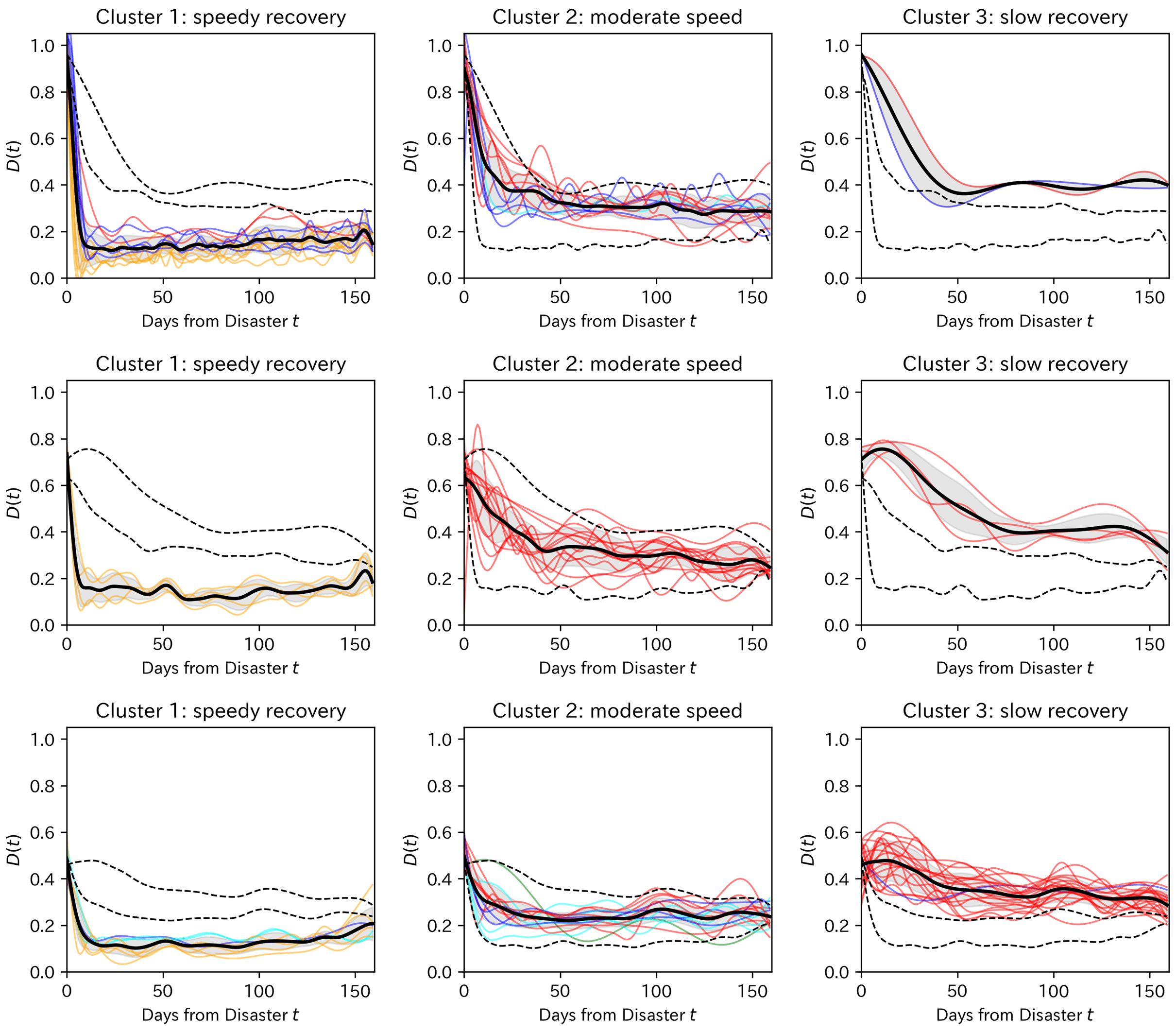}
\caption{\textbf{Clustering results of community recovery patterns.} Clusters of speedy, moderate, and slow recovery patterns for the three groups. Hierarchical clustering was applied to cluster the recovery patterns into three clusters for each group. For example, top left panel shows the communities with speedy recovery after high initial displacement rates and middle center panel shows the communities with moderate recovery speed after moderate initial displacement rates. Colors of recovery patterns represent the disaster. We can observe a mix of communities from different disasters especially for the top group with high initial displacement values. In cluster 2 (moderate recovery speed) after high initial displacement rates, communities from Tohoku and Puerto Rico resemble similar recovery patterns despite significantly distinct socio-economic characteristics.
}
\label{fig:clusters}
\end{minipage} 
\end{figure}

\begin{figure}[p]
\begin{minipage}[c][\textheight]{\textwidth}
\centering
\includegraphics[width=0.9\textwidth]{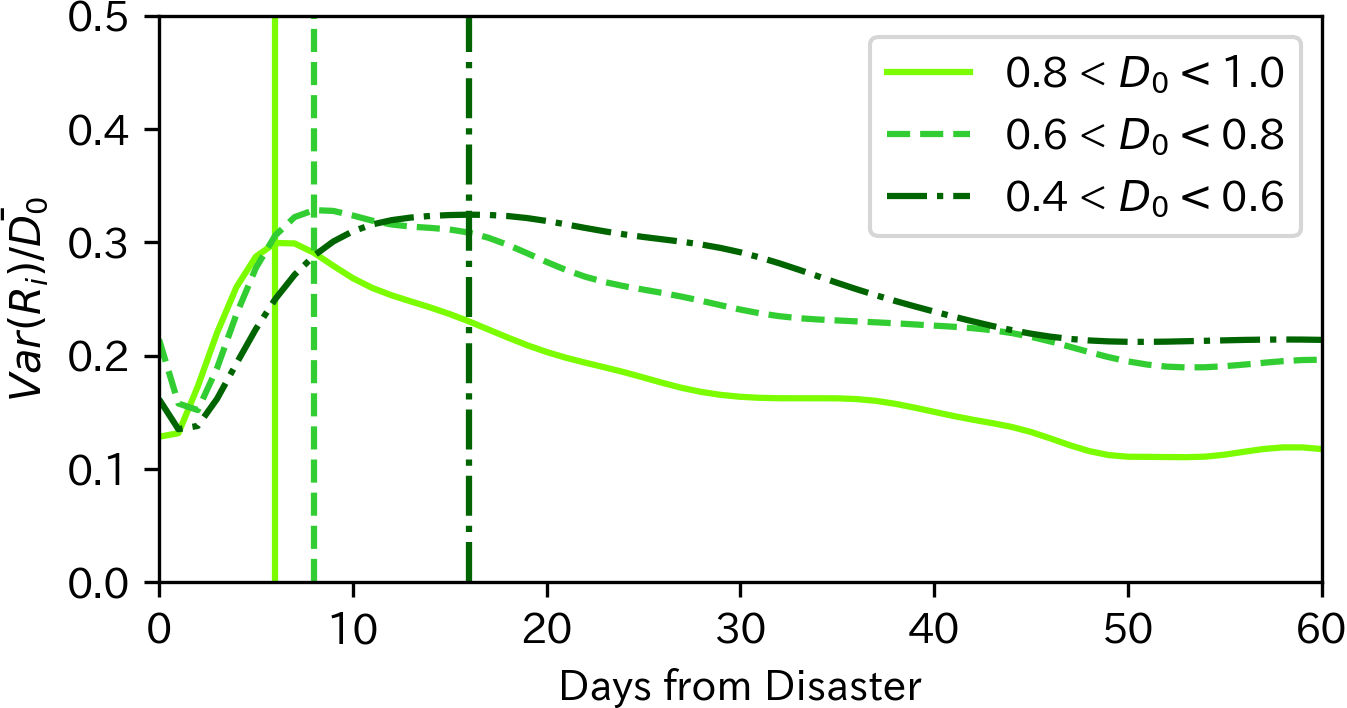}
\caption{\textbf{Temporal variance across communities.} Normalized variance of the recovery trajectories for the three groups. Y-axis shows the variance across all community recovery patterns divided by the average recovery pattern for each group. We can observe the differences in timing of maximum variance across the three groups. For the group with highest initial displacement values, the timing of maximum variance was earlier than the group with smaller initial displacement values. This implies that in communities with smaller initial displacement rates, people decide whether or not to evacuate by observing the short term (1 week) damage status of the community (e.g. power outages and water shortages) after the disaster.
}
\label{fig:variance}
\end{minipage}
\end{figure}

\begin{table}
\begin{minipage}[c][\textheight]{\textwidth}
\centering
\caption{\textbf{Statistics of GPS data for all disasters.} For all disasters, GPS location data of affected individuals were observed for 6 months before and after the disaster. All datasets had more than 30 datapoints per day for each individual on average, allowing us to accurately track where each individual stayed every night after the disaster.}
\begin{tabular}{l|rrrr}
\toprule
Disaster Name & Area of Study &  \begin{tabular}{c} Observed \\ Period \end{tabular} & \begin{tabular}{r} Affected \\ Users \end{tabular} & \begin{tabular}{r} Observations \\(/day/user) \end{tabular} \\
\midrule
Hurricane Maria & Puerto Rico & \begin{tabular}{c} 2017/9/15- \\ 2018/3/15 \end{tabular} & 53,511  & 82.8\\[0.5cm]
Hurricane Irma & Florida, USA & \begin{tabular}{c} 2017/9/1- \\ 2018/3/1 \end{tabular} & 173,0326  &  97.0\\[0.5cm]
Tohoku Tsunami & Tohoku, Japan & \begin{tabular}{c} 2011/3/1- \\ 2011/9/1 \end{tabular} & 68,416 & 33.4 \\[0.5cm]
Kumamoto Earthquake & Kumamoto, Japan & \begin{tabular}{c} 2016/4/1- \\ 2016/10/1 \end{tabular} & 80,933 & 40.7\\[0.5cm]
Kinugawa Flood & Ibaraki, Japan & \begin{tabular}{c} 2015/9/1- \\ 2016/3/1 \end{tabular} & 2,580 & 46.0\\
\bottomrule
\end{tabular}
\label{aboutGPS}
\end{minipage}
\end{table}

\begin{table}[p]
\begin{minipage}[c][\textheight]{\textwidth}
\centering

\begin{tabular}{l|lllll}
\toprule
\textbf{Tohoku Tsunami} & $r_h$ & $N_h$  & $MI$ & $t_r$ & $d_p$ \\
\midrule
Housing Damage Rate $r_h$ & 1 & & & & \\
Number of Households $N_h$ & -0.253   & 1 & & & \\
Median Income  $MI$ & -0.175   & 0.377  & 1 & & \\
Proximity to Cities $d_p$ & 0.217 & -0.208 & 0.389 & 1 & \\
Power Recovery Time $t_r$ & 0.702 & -0.131 & -0.029 & -0.064 & 1 \\
\bottomrule
\end{tabular}

\vspace*{0.3 cm}

\begin{tabular}{l|lllll}
\toprule
\textbf{Hurricane Irma} & $r_h$ & $N_h$  & $MI$ & $t_r$ & $d_p$ \\
\midrule
Housing Damage Rate $r_h$ & 1 & & & & \\
Number of Households $N_h$ & 0.02   & 1 & & & \\
Median Income  $MI$ & -0.030   & 0.307  & 1 & & \\
Proximity to Cities $d_p$ & -0.058& -0.036 & -0.260 & 1 & \\
Power Recovery Time $t_r$ & 0.725 & -0.089 & -0.093 & -0.077 & 1 \\
\bottomrule
\end{tabular}

\vspace*{0.3 cm}

\begin{tabular}{l|lllll}
\toprule
\textbf{Hurricane Maria} & $r_h$ & $N_h$  & $MI$ & $t_r$ & $d_p$ \\
\midrule
Housing Damage Rate $r_h$ & 1 & & & & \\
Number of Households $N_h$ & -0.306   & 1 & & & \\
Median Income  $MI$ & -0.266   & 0.413  & 1 & & \\
Proximity to Cities $d_p$ & 0.282& -0.322 & -0.024 & 1 & \\
Power Recovery Time $t_r$ & 0.190 & 0.107 & -0.027 & -0.180 & 1 \\
\bottomrule
\end{tabular}

\vspace*{0.3 cm}

\begin{tabular}{l|lllll}
\toprule
\textbf{Kumamoto Earthquake} & $r_h$ & $N_h$  & $MI$ & $t_r$ & $d_p$ \\
\midrule
Housing Damage Rate $r_h$ & 1 & & & & \\
Number of Households $N_h$ & -0.166   & 1 & & & \\
Median Income $MI$ & 0.055   & 0.616  & 1 & & \\
Power Recovery Time $t_r$ & 0.530 & -0.267 & -0.111 & 1 &\\
Proximity to Cities $d_p$ & 0.478 & -0.087 & -0.082 & 0.064 & 1               \\
\bottomrule
\end{tabular}

\caption{\textbf{Correlation between the variables used for regression.}}
\label{corr}
\end{minipage}
\end{table}

\begin{table}[p]
\begin{minipage}[c][\textheight]{\textwidth}
\centering
\begin{tabular}{l|cccc}
\toprule
& \multicolumn{4}{c}{\textbf{Dependent Variable: $D_{0}$}} \\ 
& Tohoku Tsunami & Hurricane Irma &  Hurricane Maria & Kumamoto Eq.  \\
\midrule
Intercept 
& 
\begin{tabular}{c} 0.13 \\ (0.12) \end{tabular} & \begin{tabular}{c} 0.16*** \\ (0.04) \\ \end{tabular} & \begin{tabular}{c} 0.76*** \\ (0.06) \end{tabular}&  \begin{tabular}{c} 0.006 \\ (0.016) \end{tabular}
\\[0.4cm]
Housing Damage Rate      
& 
\begin{tabular}{c} 0.97*** \\ (0.15) \end{tabular}& \begin{tabular}{c} 0.62*** \\ (0.11) \end{tabular} & \begin{tabular}{c} 0.06 \\ (0.10) \end{tabular} &  \begin{tabular}{c} 0.33*** \\ (0.04) \end{tabular} 
\\[0.4cm]
Number of Households          
& 
\begin{tabular}{c} -0.15 \\ (0.21) \end{tabular}& \begin{tabular}{c} -0.09 \\(0.08) \end{tabular}& \begin{tabular}{c} -0.33* \\ (0.19) \end{tabular} &  \begin{tabular}{c} 0.12*** \\ (0.03) \end{tabular} 
\\[0.4cm]
Median Income           
& 
\begin{tabular}{c} 0.25 \\ (0.22) \end{tabular}& \begin{tabular}{c} 0.04 \\ (0.07) \end{tabular}& \begin{tabular}{c} -0.40*** \\ (0.13) \end{tabular} &  \begin{tabular}{c} 0.10*** \\ (0.03) \end{tabular} 
\\[0.4cm]
Proximity to Cities    
& 
\begin{tabular}{c} -0.39* \\ (0.19) \end{tabular}& \begin{tabular}{c} 0.072 \\ (0.08) \end{tabular}& \begin{tabular}{c} -0.18* \\ (0.09)  \end{tabular} &  \begin{tabular}{c} 0.014 \\ (0.038) \end{tabular} \\
\midrule
Observations & 31 & 49 & 78 & 50 \\
$R^2$ & 0.63 & 0.45 & 0.26 & 0.78 \\
Significance $F$ & $< \,$0.001*** & $< \,$0.001*** & $< \,$0.001*** & $< \,$0.001*** \\
\bottomrule
\end{tabular}
\caption{\textbf{Regression results for initial displacement rates.} }
\label{initialreg}
\end{minipage}
\end{table}

\begin{table}[p]
\begin{minipage}[c][\textheight]{\textwidth}
\centering
\begin{tabular}{l|cccc}
\toprule
& \multicolumn{4}{c}{\textbf{Dependent Variable: $D_{160}$}} \\ 
& Tohoku Tsunami & Hurricane Irma &  Hurricane Maria & Kumamoto Eq.  \\
\midrule
Intercept 
& 
\begin{tabular}{c} 0.20*** \\ (0.04) \end{tabular} & \begin{tabular}{c} 0.11*** \\ (0.02) \\ \end{tabular} & \begin{tabular}{c} 0.25*** \\ (0.03) \end{tabular}&  \begin{tabular}{c} 0.06*** \\ (0.02) \end{tabular}
\\[0.4cm]
Housing Damage Rate      
& 
\begin{tabular}{c} 0.33*** \\ (0.09) \end{tabular}& \begin{tabular}{c} 0.39*** \\ (0.08) \end{tabular} & \begin{tabular}{c} 0.09** \\ (0.04) \end{tabular} &  \begin{tabular}{c} 0.09* \\ (0.05) \end{tabular} 
\\[0.4cm]
Number of Households          
& 
\begin{tabular}{c} 0.02 \\ (0.08) \end{tabular}& \begin{tabular}{c} -0.13*** \\(0.04) \end{tabular}& \begin{tabular}{c} -0.02 \\ (0.06) \end{tabular} &  \begin{tabular}{c} 0.03 \\ (0.03) \end{tabular} 
\\[0.4cm]
Median Income           
& 
\begin{tabular}{c} 0.005 \\ (0.09) \end{tabular}& \begin{tabular}{c} 0.05 \\ (0.03) \end{tabular}& \begin{tabular}{c} 0.11** \\ (0.05) \end{tabular} & \begin{tabular}{c} 0.03 \\ (0.03) \end{tabular} 
\\[0.4cm]
Proximity to Cities    
& 
\begin{tabular}{c} -0.05 \\ (0.08) \end{tabular}& \begin{tabular}{c} 0.11** \\ (0.04) \end{tabular}& \begin{tabular}{c} -0.01 \\ (0.04)  \end{tabular} & \begin{tabular}{c} -0.001 \\ (0.04) \end{tabular} 
\\[0.4cm]
Infrastructure Recovery    
& 
\begin{tabular}{c} -0.19* \\ (0.09) \end{tabular}& \begin{tabular}{c} 0.58*** \\ (0.21) \end{tabular}& \begin{tabular}{c} -0.05 \\ (0.05)  \end{tabular} & \begin{tabular}{c} 0.045 \\ (0.035) \end{tabular} 
\\
\midrule
Observations & 31 & 49 & 78 & 50 \\
$R^2$ & 0.42 & 0.61 & 0.14 & 0.28 \\
Significance $F$ & 0.013** & $< \,$0.001*** & 0.057* & 0.010** \\
\bottomrule
\end{tabular}
\caption{\textbf{Regression results for long term displacement rates.}}
\label{longtermreg}
\end{minipage}
\end{table}

\begin{table}[p]
\begin{minipage}[c][\textheight]{\textwidth}
    \centering
    \begin{tabular}{l|ccc}
\toprule
& \multicolumn{3}{c}{Dependent Variables: $D_{LT}$} \\
 & \multicolumn{1}{c}{$ 0.8 < D_{0} \le 1.0$} & \multicolumn{1}{c}{$ 0.6 < D_{0} \le 0.8$} & \multicolumn{1}{c}{$ 0.4 < D_{0} \le 0.6$} \\
\midrule
Intercept & \begin{tabular}{c} 0.35*** \\ (0.077) \end{tabular} & \begin{tabular}{c} 0.47***  \\ (0.15) \end{tabular} & \begin{tabular}{c} 0.35***  \\ (0.050) \end{tabular} \\[0.4cm]
Housing Damage Rate $r_h$ & \begin{tabular}{c} 0.72*** \\  (0.10) \end{tabular} & \begin{tabular}{c} 0.43**  \\ (0.19) \end{tabular}  & \begin{tabular}{c} 0.20**  \\  (0.08) \end{tabular}   \\[0.4cm]
Number of Households $N_h$ & \begin{tabular}{c} -0.24* \\  (0.12) \end{tabular} & \begin{tabular}{c} -0.18 \\  (0.16) \end{tabular} &  \begin{tabular}{c} 0.030  \\ (0.105) \end{tabular} \\[0.4cm]
Median Income $MI$ & \begin{tabular}{c} -0.48***  \\ (0.12) \end{tabular} & \begin{tabular}{c} -0.21*  \\ (0.11) \end{tabular} & \begin{tabular}{c} -0.30**  \\ (0.082)
\end{tabular} \\[0.4cm]
Proximity to Cities $d_{p}$ & \begin{tabular}{c} -0.45***  \\ (0.15) \end{tabular} &  \begin{tabular}{c} 0.02  \\ (0.13) \end{tabular} & \begin{tabular}{c} 0.18  \\ (0.12) \end{tabular} \\
\midrule
Number of Observations & 29 & 24 & 48 \\
$R^2$ & \multicolumn{1}{c}{0.74} & \multicolumn{1}{c}{0.58} & \multicolumn{1}{c}{0.63} \\
Significance $F$ & $< \,$0.001*** & $< \,$0.001*** & $< \,$0.001*** \\
\bottomrule
\end{tabular}
    \caption{\textbf{Regression results for long term displacement rates for groups with different initial displacement rates.}
    }
    \label{tab:my_label}
    \end{minipage}
\end{table}

\end{document}